\documentclass[12pt,oneside]{amsart}

\usepackage{amsmath}               
\usepackage{amsthm}                
\usepackage{times}
\usepackage{hyperref}

\theoremstyle{plain}
\theoremstyle{plain}
\theoremstyle{plain}
\theoremstyle{plain}
\theoremstyle{definition}
\theoremstyle{remark}

\numberwithin{equation}{section}

\title[Comment on {\tt{gr-qc/0303043}}]%
{Comment on "Spherically symmetric perfect fluid in area-radial coordinates" by Iguchi \emph{et al}}

\author[R.\ Giamb\`o ,\ F.\ Giannoni]{Roberto Giamb\`o, Fabio Giannoni}
\email{roberto.giambo@unicam.it, fabio.giannoni@unicam.it}
\author[G.\ Magli]{Giulio Magli}
\email{magli@mate.polimi.it}
\author[P.\ Piccione]{Paolo Piccione}
\email{paolo.piccione@unicam.it, (piccione@ime.usp.br)}

\begin{document}

\begin{abstract}
In this short note we comment about some criticisms -- appeared in a recent paper by Iguchi \emph{et al} -- to our previous works on gravitational collapse of perfect fluids. We show that those criticisms are incorrect on their own.
\end{abstract}

\maketitle


Gravitational collapse of perfect fluid is a widely treated topic, especially with numerical simulations. To perform an analytical study, studying the line element in comoving coordinates turns out to be a hard task, and the  so--called area--radius coordinates reveal extremely useful for this purpose, though on the other side less intuitive for understanding the underlying physical situation. In the following we are going to comment about some criticisms to our work \cite{grg,cqg} appeared in a recent post \cite{har} by Iguchi \emph{et al}.

Dealing with a perfect fluid with linear equation of state $p=\alpha\varepsilon$ (let us take $\alpha\in(0,1)$), yields the following expression for the matter density $\rho$:
\[
\rho^{\alpha+1}=-\frac{\Psi_{,R}}{4\pi\alpha R^2},
\]
where $\Psi(r,R)$ is the Misner--Sharpe mass function.
In our works we basically rely on an analytical form for the mass function $\Psi(r,R)$, that we assume to be Taylor expandable at $(0,0)$:
\[
\Psi(r,R)=\frac h2 \left(r^3 -\frac{\alpha}{\alpha+1}
R^3\right)+\sum_{i+j=3+k} \Psi_{ij}r^i R^j+\ldots,
\]
where the coefficients of third order are found out from the relation, involving $R'$ -- the  derivative of $R$ with respect to the \emph{comoving} variable $r$ --
\[
R'=-\frac{\alpha}{\alpha+1}\frac{\Psi_{,r}}{\Psi_{,R}},
\]
using the initial position $R=r$ at the first comoving instant $t=0$. 

Iguchi \emph{et al} claim that, with such an expansion for the mass term, the central density is constant. But of course one has
\[
\frac{\Psi_{,R}}{R^2}=-\frac {3h}2\frac{\alpha}{\alpha+1}+\sum_{i+3=3+k}j\Psi_{ij}r^i R^{j-3}+\ldots,
\]
and  the sum above may contain, for instance, terms like $r^i/R^2$, which are obviously not regular passing to the limit as $(r,R)$ goes to $(0,0)$. The above expression, moreover, shows that the energy of non central shells (i.e. fixing $r=r_0>0$) does diverge, if terms with negative power of $R$ are present. 

On the other side, one may argue that, fixing $r=0$ above, terms with negative power of $R$  disappear. But this line of reasoning doesn't show to be correct, since $r=0$ axes of the $(r,R)$ plane is not a physically meaningful direction in computing the limit of any quantity as $(r,R)$ goes to $(0,0)$.
Indeed, passing from comoving system $(t,r)$ to $(r,R)$ plane, the condition $R(t,0)=0$ maps the set $\{(t,0)\}$ into a single point -- the origin of $(r,R)$ system, namely. Fixed a shell $r_0$, the behavior of some quantity (energy density, say) can be studied for any $R\in[0,r_0]$. Therefore,
the only point in the central shell $r=0$ which makes physically sense is $R=0$,
where obviously all quantities considered in $(r,R)$ coordinates have no meaning. In short, all the relations involving coordinate changes are valid only where this coordinate change in regular - then $r>0$ in our case --
and area--radius system origin $(r=0,R=0)$ cannot be considered either a regular or a
singular point. 

Again, Iguchi \emph{et al} claim that the equation for $R'$ above implies, with an analytical mass function, that $R'$ goes like $(r/R)^2$ for all comoving time $t$. One can see, instead, that the limit to $(0,0)$ depends on the function $\Psi_{,R}/R^2$, whose expression we derived above, and which is not regular near the origin. By the way, one can easily check that also marginally bound dust solutions in area--radius coordinates leads to a similar paradox: indeed, in this case it is
\[
R'=\frac{\Psi_{,r}}{3\Psi}\left[R-r\left(\frac rR\right)^{1/2}\right]+\left(\frac rR\right)^{1/2}
\]
and since $\frac{\Psi_{,r}}{3\Psi}\cong\frac1r$ one finds $R'\cong \left(\frac Rr\right)$, from which one should in principle argue that $R\cong r$. This apparently paradoxical results (which is actually true only up to central singularity formation, i.e. $\forall t<t_s(0)$, as we well know from the analysis in comoving coordinates) again comes out since $(r,R)$--system is not able to make a distinction between the singular and the regular central shell.

Actually area-radius coordinates, though may be of great help in carrying out some of 
the calculations, are not helpful to determine whether the central shell
gets singular or not. For this aim, comoving coordinates must be involved
again to prove that the limit of the singular curve $t_s(r)$ is finite (and
nonzero of course) as $r\to 0$.

\end{document}